\documentclass[aps,reprint]{revtex4-2}

\usepackage{dcolumn}
\usepackage{bm}
\usepackage{amsthm,amsmath,amssymb,color,mathptmx,times}
\usepackage{graphicx,epstopdf}
\usepackage{blindtext}

\newcommand{\edit}[1]{#1}

\begin{document}
	
	\title{Dragon kings in self-organized criticality systems}
	\author{Guram Mikaberidze}
	\affiliation{Department of Mathematics, University  of  California, Davis, CA, 95616, USA}
	\author{Arthur Plaud}
	\affiliation{Département de Physique de l'Ecole Normale Supérieure, 24 rue Lhomond, 75231 Paris Cedex 05, France}
	\author{Raissa M. D'Souza} 
	\affiliation{University  of  California, Davis, CA, 95616, USA}
	\affiliation{Santa Fe Institute, Santa Fe, NM, 87501, USA}
	\date{\today}
	
	\begin{abstract}
		The spontaneous emergence of scale invariance, called self-organized criticality (SOC), is often attributed to a second-order absorbing-state phase transition (ASPT). Many real-world systems display SOC, yet extreme events are often overrepresented, causing significant disruption, and are called dragon kings (DK). We show analytically that the tradeoff between driving impulse and dissipation rate can create DKs in a second-order ASPT. This establishes that DKs exist in SOC systems, reveals a taxonomy of DKs, and shows that larger dissipation and smoother driving lower risk of extreme events.
	\end{abstract}
	\maketitle
	
	\emph{Introduction:}
	Many natural and engineered systems exhibit properties that follow a broad-scale distribution. For power-law distributions this is often explained through the framework of self-organized criticality (SOC). Examples are numerous and diverse, from electric power grids \cite{dobson2007complex,dsouza2012suppressing} and social media networks \cite{gleeson2014competition,dmitriev2021identification} to solar flares \cite{nishizuka2009power,aschwanden201625}, possibly brain dynamics \cite{beggs2003neuronal,fontenele2019criticality} and even the multiverse \cite{kartvelishvili2021self}. The ubiquity of power-law distributions in nature motivated the now seminal Bak-Tang-Wiesenfeld (BTW) sandpile model \cite{BTW1987}. This model sparked broad and intense research of SOC, which converged on the second-order absorbing-state phase transition (ASPT) mechanism \cite{dickman1998absorbingSOC} as a widely accepted explanation of the phenomenon \cite{pruessner2012self, zapperi2022crackling}.
	
\begin{figure}[b]
		\includegraphics[width=1\linewidth]{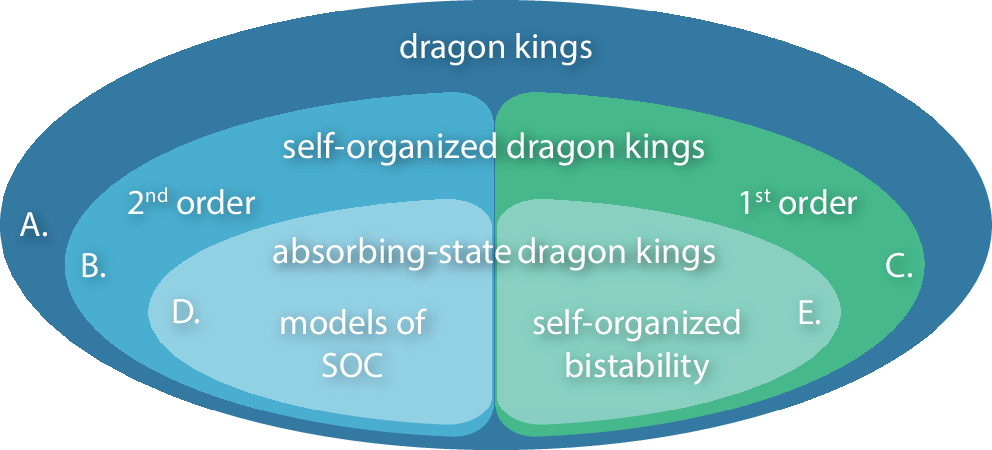}
		\caption{Dragon kings have been observed in many scenarios, we organize them by their underlying nature. \edit{The inner sets denoted as self-organized dragon kings rely on dissipation $\epsilon$ and driving $\Delta E$.}}
		\label{classification} 
	\end{figure}

	Often, in real systems, smaller events follow a power-law distribution, yet the extreme events violate the power-law paradigm by being significantly larger and overrepresented. This creates a peak at the tail of the distribution. Sornette dubbed such events dragon kings (DK) \cite{sornette2009dragon_kings,sornette2012dragon}. Dragon - to stress their unique origins, distinguishing them from smaller events, and king - to underline their disproportionate impact. DKs are usually generated by an endogenous, self-amplifying mechanism, different from mechanisms driving smaller events. They have been observed in various contexts including drawdowns in financial markets \cite{sornette2015financial_dk}, nuclear reactor leaks \cite{sornette2017nuclear_dk}, city sizes \cite{sornette2012city_dk}, marine particle sizes \cite{bochdansky2016deep_sea}, neuronal activity during epileptic seizures \cite{de2012dungeons, mishra2018dragon}, thermoacoustics experiments \cite{premraj2021catastrophic_transition},
	earthquakes \cite{sachs2012examples_and_soc}, etc. Despite their prevalence and impact, the analytic study of DKs is still quite limited. 
	Here we establish a second order ASPT mechanism for DKs, showing that DKs can exist in SOC systems and revealing a taxonomy of DK systems (Fig. \ref{classification}).
	
	Enigmatic ``peaks," ``bumps," or ``humps" have been observed in previous SOC studies \cite{grassberger1993forest_fire, zapperi1995sobp, amaral1996self, pruessner2012self, sachs2012examples_and_soc, watanabe2015fractal, dsouza2018sodk, kinouchi2019stochastic}. In special cases, the bump can be explained by the accumulation of events capped by finite system-size \cite{grassberger1993forest_fire, pruessner2012self, cavalcante2013predictability}. However, this mechanism is only relevant for the distributions with a trivial power-law exponent $\tau=1$ \cite{pruessner2012self}, and even then, the bump could be evenly spread across the whole distribution. In other cases, a bump can result from a finite driving rate \cite{queiroz2001barkhausen, hwa1992running, pradhan2021time}, however vanishing driving rate has long been recognized as a requirement for SOC \cite{vespignani1998soc_mft, dickman2000paths}. We reconcile bumps in SOC by showing that the standard SOC framework surprisingly produces DK events through the interplay of dissipation and an intrinsic quantity that we call the driving impulse, which has received only limited attention \cite{cafiero1995local, munoz2009nonconservative}. Furthermore, we derive a new necessary condition for SOC.
	
	The self-organization (SO) mechanism, tunes a system to balance the tradeoffs between two external ``forces"  \cite{pruessner2012self, dsouza2018sodk}.
	The balance usually occurs near a phase transition point between active and inactive phases. The first ``force" is driving  (e.g., sand deposition in BTW sandpile model \cite{BTW1987}), which increases the control parameter $E$ (e.g., the particle density)  at a slow constant rate $h$. The second ``force" is dissipation $\epsilon$, reducing $E$ proportional to the current activity level (e.g., boundary or bulk dissipation of sand). 
	
\begin{figure}[t]
		\includegraphics[width=\linewidth]{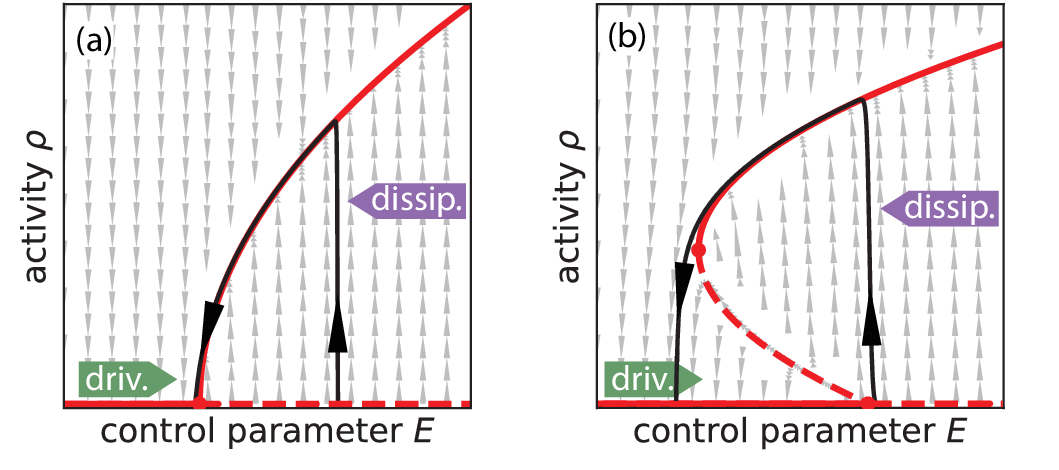}
		\caption{Self-organization around: (a) a second-order ASPT, creating SOC or DKs depending on regime; (b) a first-order ASPT creating hysteresis with DKs. Solid (dashed) lines indicate stable (unstable) steady-states of the underlying phase transition model without SO (with $E$ fixed, often called ``static"). Black trajectories are examples of possible DK events.}
		\label{phase_plots}
	\end{figure}

	Self-organization around a second-order ASPT is the accepted theory of SOC (Fig. \ref{phase_plots}a). The order parameter is the stable stationary activity. When the system is subcritical ($E<E_c$), the steady state is inactive. Thus, the SO mechanism causes little dissipation and drives the system toward $E=E_c$. In contrast, the supercritical system ($E>E_c$) has an active, stable, stationary state in the thermodynamic limit. Thus, the SO mechanism dissipates $E$ in a cascade of activity, driving the system back to $E=E_c$. The system becomes balanced at the critical point producing power-law distributions in response functions. The size of the cascades follows the distribution $P(S)=S^{-\tau} G(S/S_\epsilon)$ in the limit of large system size $N\rightarrow\infty$ (thermodynamic limit), small dissipation probability $\epsilon\rightarrow 0$ (conservative limit) and vanishing ratio of driving rate to dissipation $h/\epsilon\rightarrow 0$ (adiabatic limit) \cite{vespignani1998soc_mft,pruessner2012self, noel2013controlling,zapperi2022crackling}. Here $S_\epsilon$ is the cutoff due to the dissipation $\epsilon$ (note, $S_\epsilon\sim \epsilon^{-1/\sigma}\rightarrow\infty$ as $\epsilon\rightarrow 0$).
	
	If instead of a second-order ASPT, one considers a first-order ASPT as shown in Fig. \ref{phase_plots}b, the SO mechanism leads to hysteresis. Mu{\~n}oz and collaborators established this in recent papers \cite{munoz2016so_bistability, munoz2020sob_brain}  (see also \cite{sornette1996landau-ginsburg-soc}) and dubbed this phenomenon ``self-organized bistability". The corresponding cascades follow a power-law distribution with a DK bump (set E in Fig. \ref{classification}). This is a step toward unifying DKs and SOC, but SOC is described by a second-order ASPT.
	
	For a first-order ASPT, the emergence of DKs is illustrated in Fig. \ref{phase_plots}b. Plotted is the activity of the system $\rho$ against control parameter $E$. For small values of $E$ the system is inactive ($\rho=0$). Driving increases $E$, but since the inactive branch is stable, small fluctuations can only start minor cascades, causing little dissipation. Once driven into the unstable region, a DK event begins (black curve in Fig. \ref{phase_plots}b). The system is repelled from the newly unstable quiescent branch, while attracted to the stable active branch. Repulsion increases linearly with distance from the unstable branch, and the activity grows exponentially, leading to a self-amplifying DK mechanism. Once on the active branch, the macroscopic activity causes dissipation, reducing $E$ until the saddle-node bifurcation point after which the system falls back into a quiescent steady state, concluding the single DK event. Then the cycle repeats. Examples of such first-order absorbing state DKs include the facilitated sandpile introduced in \cite{munoz2016so_bistability}, and more recently, the BTW-Kuramoto model \cite{mikaberidze2022sandpile}.

	\emph{Dragon kings in models of SOC:}
	We next show how DKs can arise from a second-order ASPT (set D in Fig. \ref{classification}). Examples of such DKs include the self-organized quasi-critical neuronal models \cite{ kinouchi2019stochastic} and the rice-pile model \cite{amaral1996self} (see also \cite{grassberger1993forest_fire, sachs2012examples_and_soc, watanabe2015fractal}).
	
	
	
	For a DK to be triggered in a second-order ASPT model of SOC it must be driven far into the supercritical region (Fig. \ref{phase_plots}a). Such a supercritical state is quiescent (e.g., when all nodes are below or at capacity in BTW sandpile model) and stable under some perturbations due to the discrete nature of the state-space. Stable perturbations move the system to another quiescent configuration (e.g., when a grain of sand is added to a node that was below its capacity) or displace it in the stable direction, causing only minor activity (e.g., when sand arrival causes a local cascade with only a few topplings). Thus the system advances deeper into the supercritical regime with a non-zero probability. Once the macroscopic activity is eventually triggered, it lasts until the surplus of the control parameter ($E-E_c$) gets dissipated and $E$ is reduced back to $E_c$. Intuitively, a massive event is more likely when driving increases $E$ in large increments $\Delta E$ and the dissipation $\epsilon$ is small. 
	
	SOC requires the adiabatic limit where the driving rate $h\rightarrow 0$, allowing all cascading activity to die down before performing the next driving step. Non-zero $h$ creates overlapping cascades, distorting the power-law and potentially creating a DK peak \cite{white2003driving,pradhan2021time,zapperi2022crackling}. The adiabatic limit is realized ideally in computer simulations where the code ``babysits" \cite{dickman2000paths} the dynamics. This introduces an infinite separation of timescales between cascading and driving dynamics,
	implying driving happens infinitely rarely on the cascading timescale, thus $h=0$. However, even with this infinite separation, the driving is still performed in discrete steps, imposing upon us another small parameter, the driving impulse $\Delta E$. 
	
	\emph{The role of the driving impulse $\Delta E$:}
	The driving impulse $\Delta E$ is an important parameter (e.g., see \cite{cafiero1995local, munoz2009nonconservative}) distinct from the driving rate $h$. To achieve SOC, one needs the stronger limit $\Delta E\rightarrow 0$ in place of $h\rightarrow 0$. This limit is usually implicit in SOC models. For example, in the BTW sandpile model, the system is driven by adding one grain of sand at a time, $\Delta E=\frac{1}{N}$ which automatically vanishes in the thermodynamic limit. We will show that even in the dual limit $\epsilon,\Delta E\rightarrow0$, the interrelation of $\epsilon$ and $\Delta E$ determines the presence or absence of DK events in otherwise power-law distributions. In contrast, for $\Delta E$ sufficiently large, all avalanches are system-wide \cite{cafiero1995local}. 
	
	\emph{DK condition in models of SOC:}
	Consider a second-order ASPT (Fig. \ref{phase_plots}a). Without driving and dissipation, perturbations introduced in a quiescent state will die down with probability $p$ or grow into macroscopic activity with probability $1-p$. Clearly, $p=1$ in the subcritical states and $p<1$ in supercritical states. Therefore $p$ as a function of $E$ undergoes a phase transition at $E=E_c$. Assuming that near criticality $p$ follows scaling behavior standard for continuous phase transitions \cite{pathria2016statistical}, we write
	\begin{equation}\label{p(E)}
		p(E) = \begin{cases}
			1											& \text{for } E\le E_c \\
			1-k(E-E_c)^\lambda 		& \text{for } E>E_c.
		\end{cases}
	\end{equation}
	The critical exponent, $\lambda$, depends on the universality class of the system. The constants $k$ and $E_c$ depend on the microscopic rules of the model.
	
	Let us consider self-organization with a small driving impulse and vanishing dissipation around such a second-order ASPT. Driving increases $E$ by a constant amount $\Delta E$ at every iteration which simultaneously introduces a small perturbation to allow a cascade to begin \footnote{Note that driving can be independent of perturbation. In the self-organized branching process, \cite{zapperi1995sobp} the driving increments the branching probability while perturbation is introduced by activating a single site.}. In the limit of vanishing dissipation $\epsilon\rightarrow 0$, minor cascades do not impact the control parameter, and the system moves deeper into the supercritical phase. Eventually, a macroscopic cascade will get triggered pushing the system to the active stable branch. Dissipation then reduces $E$ proportional to the current level of activity $\rho$, and the system evolves along the active branch lowering $E$ until it reaches $E_c$ and falls into an absorbing state. This is a single DK cascade (see the black curve in Fig. \ref{phase_plots}a).
	
	The macroscopic cascade starts with supercritical $E>E_c$ and terminates at $E_c$. The size of the cascade $S$ causes dissipation $\epsilon S$ which means $(E-E_c)=\epsilon S$ or
	\begin{equation}
		S(E)=\frac{E-E_c}{\epsilon}.
	\end{equation}
	
	We want to find the average size of a DK event in the limit $\epsilon\rightarrow 0$. Starting at $E_c$, the control parameter increases by $\Delta E$ on every iteration until a DK eventually happens. After $j$ uneventful iterations, $E_j=E_c+j\Delta E$. The probability that a DK will get triggered after exactly $j$ iterations and not earlier is given by $q_j=(1-p(E_j))\prod_{i=0}^{j-1} p(E_i)$. Thus the expected size of a DK is
	\begin{equation}\label{s_DK_initial}
		\begin{split}
			S_{DK} &= \sum_{j=0}^\infty S(E_j)q_j \\
			&= \sum_{j=0}^\infty     \frac{j\Delta E}{\epsilon}    (1-p(E_c+j\Delta E))\prod_{i=0}^{j-1} p(E_c+i\Delta E).
		\end{split}
	\end{equation}
	We can simplify this using Taylor expansions and by approximating the sums by integrals (see Supplemental Information)
	\begin{equation}\label{s_DK}
		S_{DK} \approx \frac{1}{\epsilon}\left( \frac{1+\lambda}{k}\Delta E\right)^{\frac{1}{1+\lambda}}\Gamma\left( \frac{2+\lambda}{1+\lambda} \right).
	\end{equation}
	
	This derivation holds for small fixed $\Delta E$ in the limit $\epsilon\rightarrow 0$. In the reversed regime, with small fixed $\epsilon$ and vanishing $\Delta E\rightarrow 0$, we expect a standard SOC distribution $P(S)=S^{-\tau} G(S/S_\epsilon)$ with a sharp cutoff at the dissipative scale $S_\epsilon$. Balance between driving and dissipation determines $S_\epsilon$ \cite{vespignani1998soc_mft}
	
	
	\begin{equation}\label{dissip_balance}
		\Delta E = \sum_{S=0}^\infty \epsilon S P(S) \approx \epsilon S_\epsilon^{2-\tau} \int_0^\infty z^{1-\tau}G(z)dz
	\end{equation}
	\edit{with $\tau<2$.} We express $S_\epsilon$ as a function of $\epsilon$ and $\Delta E$
	\begin{equation}\label{s_eps}
		S_\epsilon \approx \left(\int_0^\infty z^{1-\tau}G(z)dz\right)^{-\frac{1}{2-\tau}} \left(\frac{\Delta E}{\epsilon}\right)^\frac{1}{2-\tau}.
	\end{equation}
	
	Notice that only the smaller scale between $S_\epsilon$ and $S_{DK}$ is relevant. When $S_\epsilon \ll S_{DK}$, the cascade size distribution terminates at $S_\epsilon$, and there is no DK peak. If $S_\epsilon \gg S_{DK}$, a DK peak at $S_{DK}$ violates the finite size scaling assumption $P(S)=S^{-\tau} G(S/S_\epsilon)$ and renders the dissipation cutoff $S_\epsilon$ unphysical.
	
	The competition between the two scales divides the parameter space of an SOC model into two regions, one with DK peaks and the other without. The boundary between the regions is roughly given by $S_\epsilon \sim S_{DK}$, or neglecting the coefficients in Eqs. \eqref{s_DK} and \eqref{s_eps} (see Supplemental Information for details)
	\begin{equation} \label{dk_boundary}
		\Delta E \sim \epsilon ^{\frac{(\lambda +1) (\tau -1)}{\lambda +\tau -1}}.
	\end{equation}
	Then the DK condition is:
	\begin{equation} \label{dk_cond}
		\Delta E \gg \epsilon ^{\frac{(\lambda +1) (\tau -1)}{\lambda +\tau -1}}.
	\end{equation}
	This result parallels the conjecture made by Kinouchi et al. in \cite{kinouchi2019stochastic} that there is a connection between adaptive SOC and DKs, and an earlier attempt to unify DKs with power-laws by Eliazar \cite{eliazar2017black}. It agrees with a qualitative observation by Bonachela et al. in \cite{munoz2009nonconservative} that for finite dissipation $\epsilon$, large driving impulse causes supercritical behavior with the DK bump. It also aligns with the results by Cafiero et al. that SOC requires non-zero local rigidity, which in our setting translates to $\Delta E\rightarrow0$ \cite{cafiero1995local}.  We show further that even with  $\Delta E\rightarrow0$, one can get DKs provided that $\epsilon$ scales accordingly. This provides a new necessary condition on SOC.
	
	\begin{figure}[b]
		\includegraphics[width=\linewidth]{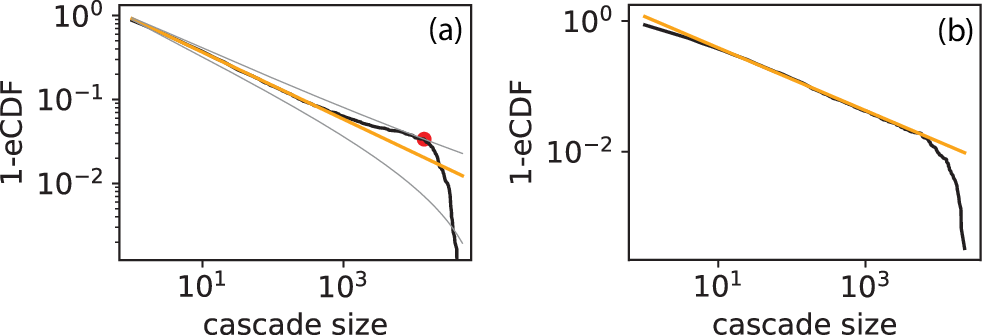}
		\caption{Illustration of DK detection based on  empirical cumulative distribution function \cite{janczura2012black}. (a) Typical distribution with DKs present \edit{(large $\Delta E$)}. The straight line is a power-law fit of the data in the second decade. The gray curves show the tightest confidence interval around this power law, which contains the most overrepresented data point. (b) Typical distribution without DKs  \edit{(small $\Delta E$)}.}
		\label{CDF}
	\end{figure}
	
	For instance, in many SOC models $\Delta E=N^{-1}$ and the effective dissipation $\epsilon$ also scales as some power of $N$ (e.g., due to open boundary conditions). The potential appearance of the DK peak in such systems will depend on these scaling laws through Eq. \eqref{dk_boundary}.

	\emph{Examples:}
	A classic example of a second-order ASPT is directed percolation. Consider a network where each node is either infected (active) or susceptible (inactive). Infected nodes infect each of their neighbors with probability $E$ and then immediately recover. Here $E$ serves as the self-organizing control parameter. A cascade is triggered by infecting a random node and lasts until all nodes recover. The infection probability $E$ increases by $\Delta E$ before every cascade (driving) and decreases by $\epsilon\rho$ at every time step during the cascade (dissipation). Here $\rho$ is the number of infected nodes (i.e., activity), and $\epsilon$ is the dissipation.
	
	This model undergoes a second-order phase transition. Under mean-field approximation, the probability of falling back into the absorbing state Eq. \eqref{p(E)} can be approximated as the extinction probability of the corresponding branching process \cite{zapperi1995sobp}. This fixes $\lambda=1$. In addition, the critical exponent for the cascade size distribution has the mean-field value $\tau=\frac{3}{2}$. Then Eq. \eqref{dk_boundary} reduces to $\Delta E \sim \epsilon^\frac{2}{3}$. Alternatively, we can write it as $\Delta E = f(\epsilon) \epsilon^\frac{2}{3}$ with $f(\epsilon)= O(1)$ which becomes a constant coefficient in the small dissipation limit
	\begin{equation}\label{dk_cond_example}
		\Delta E = c \epsilon^\frac{2}{3},
	\end{equation}
	where the constant $c=\lim_{\epsilon\rightarrow 0} f(\epsilon)$ can be determined by fitting experimental data.
	
	The same argument leads also to Eq. \eqref{dk_cond_example} for the BTW sandpile model on a network. To explore Eq. \eqref{dk_cond_example} beyond the standard fixed driving impulse $\Delta E=\frac{1}{N}$, we extend the BTW model to allow different values of $\Delta E$ by adding more than one grain during driving. The additional sand grains are deposited to non-critical sites to \edit{retain unique cascade seeds and keep activity perturbations consistant}.

	\begin{figure}[t]
		\includegraphics[width=1\linewidth]{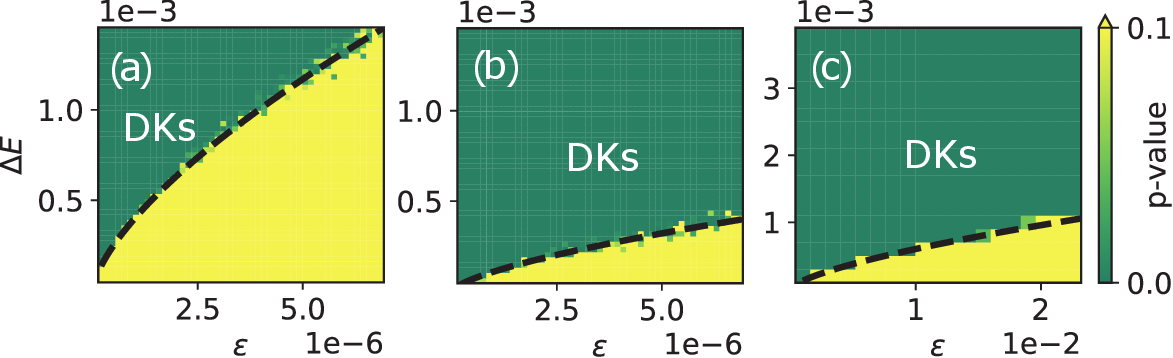}
		\caption{Parameter spaces for self-organized directed percolation (a) and (b), and for the BTW model (c), colored by the $p$-value of the statistical DK test \cite{janczura2012black}. The green (dark colored) region indicates the presence of DKs, \edit{while yellow (light colored) regions indicate power-law with a cut-off}. The black dash is the theoretical border between the two regions, Eq. \eqref{dk_cond_example} with $c$ fitted to $p\text{-value}=0.05$. (a) and (c) are for random $3$-regular networks (with $c=4$ and $c=0.013$ respectively), while (b) is for Erdős–Rényi graphs of the same density (with $c=1.1$).}
		\label{heatmap}
	\end{figure}
	
	We perform simulations for various values of dissipation $\epsilon$ and driving impulse $\Delta E$. We consider random 3-regular and Erdős–Rényi networks. Each run produces a cascade size distribution. The potential presence of a DK peak is then quantified by a $p$-value of the statistical DK detection test developed in \cite{janczura2012black}. This test considers the empirical cumulative distribution function $eCDF$ of event sizes \edit{(fraction of events larger than $S$)}. To find the $p$-value, we first fit the appropriate section of the $1-eCDF$ with a power law (see Fig. \ref{CDF}). Next, we find the most overrepresented data point in the tail, potentially a DK (marked with a dot in Fig. \ref{CDF}a). Finally, we find the tightest confidence interval accommodating this point \cite{janczura2012black}. One minus this confidence level is the $p$-value of the null hypothesis that there is no DK in the distribution.
	
	Figure \ref{heatmap} shows the parameter space of the above models colored by the $p$-values of the statistical test. The dash shows the theoretical boundary while green color (dark region) indicates the presence of DKs. Each simulation was performed on a network of $5000$ nodes for $10^5$ driving iterations. Power-laws were fitted to the cascade sizes $10<S<1000$ for directed percolation and $10<S<100$ for the BTW sandpile model. Experiments analogous to Fig. \ref{heatmap} confirm that the exponent $\frac{3}{2}$ in Eq. \eqref{dk_cond_example} is robust and independent of the degree and connectivity of the random regular and Erdős–Rényi graphs.

\begin{figure}[b]
		\includegraphics[width=1\linewidth]{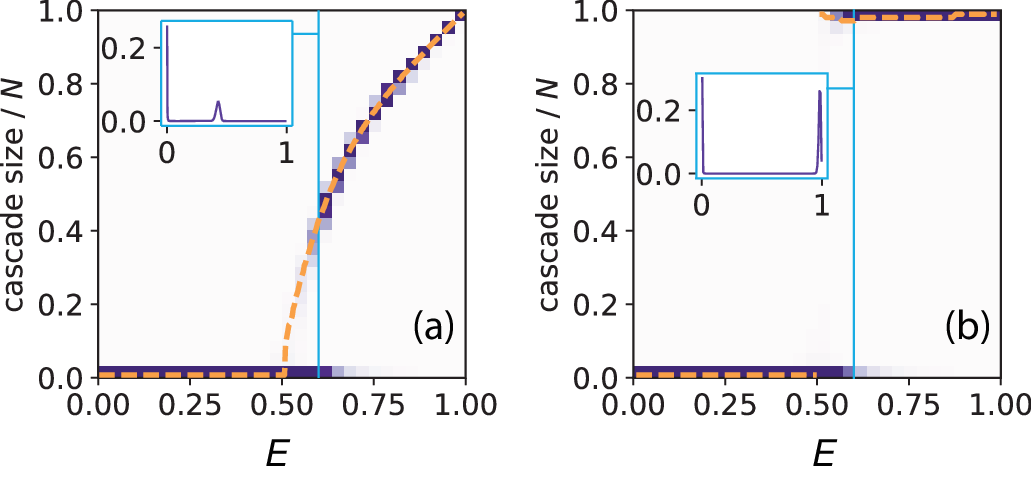}
		\caption{Histograms of cascade sizes for different values of the control parameter $E$ (weak node density) in perfectly mixed (a) inoculation and (b) complex contagion models of \cite{dsouza2018sodk}. Dark pixels indicate a high probability. The dashed line shows the largest mode (location of the right-most peak) which is: (a) continuous for inoculation model (second-order transition), and (b) discontinuous for the complex contagion models (first-order transition). Insets show vertical slices (i.e., cascade size distribution) for $E=0.6$ marked with blue (light colored) vertical line.}
		\label{sodk} 
	\end{figure}

	\emph{Dragon king taxonomy:} Self-organized DKs are not restricted to the ASPT mechanism (sets D and E in Fig. \ref{classification}). For example, the inoculation and complex contagion models described in \cite{dsouza2018sodk} self-organize around ``spreading transitions", where activity dies down in subcritical states, while in supercritical states it consists of a propagating front line \cite{cardy1985epidemic}. The cascade size distribution becomes bimodal in both models once the control parameter exceeds the critical value. The largest mode (i.e., the location of the right-most peak in the distribution) has a continuous transition in the inoculation model (Fig. \ref{sodk}a), producing second-order DKs that belong to set D of Fig. \ref{classification}. The same order parameter is discontinuous for the complex contagion model (Fig. \ref{sodk}b) producing first-order DKs which belong to set E of Fig. \ref{classification}  \footnote{Notice that the driving impulse is $\Delta E=N^{-1}$ for both models of \cite{dsouza2018sodk}. Thus the second-order DKs disappear in the thermodynamic limit unless $\epsilon$ also scales with $N$ (inoculation model). Yet, for the first-order DKs (complex contagion model), the peak remains sharply pronounced in the thermodynamic limit since it does not require the system to become strongly supercritical.}. Set B is strictly larger than D since it contains the inoculation model of \cite{dsouza2018sodk} and the self-organized branching process \cite{zapperi1995sobp}, which are not in D. Similarly, set C$\setminus$E contains the complex contagion model of \cite{dsouza2018sodk}. Set A is also larger than B$\cup$C since some DK systems do not utilize the SO mechanism \cite{odor2022powergrid, cavalcante2013predictability}.
	
	\emph{Discussion and conclusions:}
	We have shown analytically that DKs can exist in SOC systems. Even for a vanishing driving rate $h$, the competition between driving impulse $\Delta E$ and dissipation $\epsilon$ determines whether a DK peak appears in such systems. Proportionally large driving impulse $\Delta E$ can push the system deep into the supercritical phase, thus making it more susceptible to system-wide DK cascades. On the other hand, for proportionally large dissipation, small cascades are sufficient to bring the system back to the critical point, thus eliminating the DK peak from the event size distribution.
	
	
	The devastating nature of DKs is usually alleviated by their predictability, allowing preparation and planning \cite{sornette2009dragon_kings,sornette2012dragon, dsouza2018sodk, mikaberidze2022sandpile}. The analytic treatment developed in this paper could be useful for predicting and controlling DKs. The condition in Eq. \eqref{dk_cond} determines if DKs are to be expected, in which case dissipation should be increased and/or driving impulse should be decreased to avoid them. One can predict the DK probability $1-p(E)$ through Eq. \eqref{p(E)} based on the global information $E$. Alternatively, one can evaluate the risk that an ongoing cascade will become a DK by monitoring its local features \cite{dsouza2018sodk, mikaberidze2022sandpile}. E.g., the exponential growth of activity would imply the onset of the self-amplifying mechanism \cite{mikaberidze2022sandpile} (this does not require global information, but it can only give predictions after the cascade has started).
	
	\bibliographystyle{apsrev4-2}
	\bibliography{DKreferences.bib}

\begin{thebibliography}{50}%
\makeatletter
\providecommand \@ifxundefined [1]{%
 \@ifx{#1\undefined}
}%
\providecommand \@ifnum [1]{%
 \ifnum #1\expandafter \@firstoftwo
 \else \expandafter \@secondoftwo
 \fi
}%
\providecommand \@ifx [1]{%
 \ifx #1\expandafter \@firstoftwo
 \else \expandafter \@secondoftwo
 \fi
}%
\providecommand \natexlab [1]{#1}%
\providecommand \enquote  [1]{``#1''}%
\providecommand \bibnamefont  [1]{#1}%
\providecommand \bibfnamefont [1]{#1}%
\providecommand \citenamefont [1]{#1}%
\providecommand \href@noop [0]{\@secondoftwo}%
\providecommand \href [0]{\begingroup \@sanitize@url \@href}%
\providecommand \@href[1]{\@@startlink{#1}\@@href}%
\providecommand \@@href[1]{\endgroup#1\@@endlink}%
\providecommand \@sanitize@url [0]{\catcode `\\12\catcode `\$12\catcode
  `\&12\catcode `\#12\catcode `\^12\catcode `\_12\catcode `\%12\relax}%
\providecommand \@@startlink[1]{}%
\providecommand \@@endlink[0]{}%
\providecommand \url  [0]{\begingroup\@sanitize@url \@url }%
\providecommand \@url [1]{\endgroup\@href {#1}{\urlprefix }}%
\providecommand \urlprefix  [0]{URL }%
\providecommand \Eprint [0]{\href }%
\providecommand \doibase [0]{https://doi.org/}%
\providecommand \selectlanguage [0]{\@gobble}%
\providecommand \bibinfo  [0]{\@secondoftwo}%
\providecommand \bibfield  [0]{\@secondoftwo}%
\providecommand \translation [1]{[#1]}%
\providecommand \BibitemOpen [0]{}%
\providecommand \bibitemStop [0]{}%
\providecommand \bibitemNoStop [0]{.\EOS\space}%
\providecommand \EOS [0]{\spacefactor3000\relax}%
\providecommand \BibitemShut  [1]{\csname bibitem#1\endcsname}%
\let\auto@bib@innerbib\@empty
\bibitem [{\citenamefont {Dobson}\ \emph {et~al.}(2007)\citenamefont {Dobson},
  \citenamefont {Carreras}, \citenamefont {Lynch},\ and\ \citenamefont
  {Newman}}]{dobson2007complex}%
  \BibitemOpen
  \bibfield  {author} {\bibinfo {author} {\bibfnamefont {I.}~\bibnamefont
  {Dobson}}, \bibinfo {author} {\bibfnamefont {B.~A.}\ \bibnamefont
  {Carreras}}, \bibinfo {author} {\bibfnamefont {V.~E.}\ \bibnamefont
  {Lynch}},\ and\ \bibinfo {author} {\bibfnamefont {D.~E.}\ \bibnamefont
  {Newman}},\ }\href@noop {} {\bibfield  {journal} {\bibinfo  {journal} {Chaos:
  An Interdisciplinary Journal of Nonlinear Science}\ }\textbf {\bibinfo
  {volume} {17}},\ \bibinfo {pages} {026103} (\bibinfo {year}
  {2007})}\BibitemShut {NoStop}%
\bibitem [{\citenamefont {Brummitt}\ \emph {et~al.}(2012)\citenamefont
  {Brummitt}, \citenamefont {D’Souza},\ and\ \citenamefont
  {Leicht}}]{dsouza2012suppressing}%
  \BibitemOpen
  \bibfield  {author} {\bibinfo {author} {\bibfnamefont {C.~D.}\ \bibnamefont
  {Brummitt}}, \bibinfo {author} {\bibfnamefont {R.~M.}\ \bibnamefont
  {D’Souza}},\ and\ \bibinfo {author} {\bibfnamefont {E.~A.}\ \bibnamefont
  {Leicht}},\ }\href@noop {} {\bibfield  {journal} {\bibinfo  {journal}
  {Proceedings of the National Academy of Sciences}\ }\textbf {\bibinfo
  {volume} {109}},\ \bibinfo {pages} {E680} (\bibinfo {year}
  {2012})}\BibitemShut {NoStop}%
\bibitem [{\citenamefont {Gleeson}\ \emph {et~al.}(2014)\citenamefont
  {Gleeson}, \citenamefont {Ward}, \citenamefont {O’Sullivan},\ and\
  \citenamefont {Lee}}]{gleeson2014competition}%
  \BibitemOpen
  \bibfield  {author} {\bibinfo {author} {\bibfnamefont {J.~P.}\ \bibnamefont
  {Gleeson}}, \bibinfo {author} {\bibfnamefont {J.~A.}\ \bibnamefont {Ward}},
  \bibinfo {author} {\bibfnamefont {K.~P.}\ \bibnamefont {O’Sullivan}},\ and\
  \bibinfo {author} {\bibfnamefont {W.~T.}\ \bibnamefont {Lee}},\ }\href@noop
  {} {\bibfield  {journal} {\bibinfo  {journal} {Physical Review Letters}\
  }\textbf {\bibinfo {volume} {112}},\ \bibinfo {pages} {048701} (\bibinfo
  {year} {2014})}\BibitemShut {NoStop}%
\bibitem [{\citenamefont {Dmitriev}\ and\ \citenamefont
  {Dmitriev}(2021)}]{dmitriev2021identification}%
  \BibitemOpen
  \bibfield  {author} {\bibinfo {author} {\bibfnamefont {A.}~\bibnamefont
  {Dmitriev}}\ and\ \bibinfo {author} {\bibfnamefont {V.}~\bibnamefont
  {Dmitriev}},\ }\href@noop {} {\bibfield  {journal} {\bibinfo  {journal}
  {Complexity}\ }\textbf {\bibinfo {volume} {2021}} (\bibinfo {year}
  {2021})}\BibitemShut {NoStop}%
\bibitem [{\citenamefont {Nishizuka}\ \emph {et~al.}(2009)\citenamefont
  {Nishizuka}, \citenamefont {Asai}, \citenamefont {Takasaki}, \citenamefont
  {Kurokawa},\ and\ \citenamefont {Shibata}}]{nishizuka2009power}%
  \BibitemOpen
  \bibfield  {author} {\bibinfo {author} {\bibfnamefont {N.}~\bibnamefont
  {Nishizuka}}, \bibinfo {author} {\bibfnamefont {A.}~\bibnamefont {Asai}},
  \bibinfo {author} {\bibfnamefont {H.}~\bibnamefont {Takasaki}}, \bibinfo
  {author} {\bibfnamefont {H.}~\bibnamefont {Kurokawa}},\ and\ \bibinfo
  {author} {\bibfnamefont {K.}~\bibnamefont {Shibata}},\ }\href@noop {}
  {\bibfield  {journal} {\bibinfo  {journal} {The Astrophysical Journal}\
  }\textbf {\bibinfo {volume} {694}},\ \bibinfo {pages} {L74} (\bibinfo {year}
  {2009})}\BibitemShut {NoStop}%
\bibitem [{\citenamefont {Aschwanden}\ \emph {et~al.}(2016)\citenamefont
  {Aschwanden}, \citenamefont {Crosby}, \citenamefont {Dimitropoulou},
  \citenamefont {Georgoulis}, \citenamefont {Hergarten}, \citenamefont
  {McAteer}, \citenamefont {Milovanov}, \citenamefont {Mineshige},
  \citenamefont {Morales}, \citenamefont {Nishizuka} \emph
  {et~al.}}]{aschwanden201625}%
  \BibitemOpen
  \bibfield  {author} {\bibinfo {author} {\bibfnamefont {M.~J.}\ \bibnamefont
  {Aschwanden}}, \bibinfo {author} {\bibfnamefont {N.~B.}\ \bibnamefont
  {Crosby}}, \bibinfo {author} {\bibfnamefont {M.}~\bibnamefont
  {Dimitropoulou}}, \bibinfo {author} {\bibfnamefont {M.~K.}\ \bibnamefont
  {Georgoulis}}, \bibinfo {author} {\bibfnamefont {S.}~\bibnamefont
  {Hergarten}}, \bibinfo {author} {\bibfnamefont {J.}~\bibnamefont {McAteer}},
  \bibinfo {author} {\bibfnamefont {A.~V.}\ \bibnamefont {Milovanov}}, \bibinfo
  {author} {\bibfnamefont {S.}~\bibnamefont {Mineshige}}, \bibinfo {author}
  {\bibfnamefont {L.}~\bibnamefont {Morales}}, \bibinfo {author} {\bibfnamefont
  {N.}~\bibnamefont {Nishizuka}}, \emph {et~al.},\ }\href@noop {} {\bibfield
  {journal} {\bibinfo  {journal} {Space Science Reviews}\ }\textbf {\bibinfo
  {volume} {198}},\ \bibinfo {pages} {47} (\bibinfo {year} {2016})}\BibitemShut
  {NoStop}%
\bibitem [{\citenamefont {Beggs}\ and\ \citenamefont
  {Plenz}(2003)}]{beggs2003neuronal}%
  \BibitemOpen
  \bibfield  {author} {\bibinfo {author} {\bibfnamefont {J.~M.}\ \bibnamefont
  {Beggs}}\ and\ \bibinfo {author} {\bibfnamefont {D.}~\bibnamefont {Plenz}},\
  }\href@noop {} {\bibfield  {journal} {\bibinfo  {journal} {Journal of
  Neuroscience}\ }\textbf {\bibinfo {volume} {23}},\ \bibinfo {pages} {11167}
  (\bibinfo {year} {2003})}\BibitemShut {NoStop}%
\bibitem [{\citenamefont {Fontenele}\ \emph {et~al.}(2019)\citenamefont
  {Fontenele}, \citenamefont {de~Vasconcelos}, \citenamefont {Feliciano},
  \citenamefont {Aguiar}, \citenamefont {Soares-Cunha}, \citenamefont
  {Coimbra}, \citenamefont {Dalla~Porta}, \citenamefont {Ribeiro},
  \citenamefont {Rodrigues}, \citenamefont {Sousa} \emph
  {et~al.}}]{fontenele2019criticality}%
  \BibitemOpen
  \bibfield  {author} {\bibinfo {author} {\bibfnamefont {A.~J.}\ \bibnamefont
  {Fontenele}}, \bibinfo {author} {\bibfnamefont {N.~A.}\ \bibnamefont
  {de~Vasconcelos}}, \bibinfo {author} {\bibfnamefont {T.}~\bibnamefont
  {Feliciano}}, \bibinfo {author} {\bibfnamefont {L.~A.}\ \bibnamefont
  {Aguiar}}, \bibinfo {author} {\bibfnamefont {C.}~\bibnamefont
  {Soares-Cunha}}, \bibinfo {author} {\bibfnamefont {B.}~\bibnamefont
  {Coimbra}}, \bibinfo {author} {\bibfnamefont {L.}~\bibnamefont
  {Dalla~Porta}}, \bibinfo {author} {\bibfnamefont {S.}~\bibnamefont
  {Ribeiro}}, \bibinfo {author} {\bibfnamefont {A.~J.}\ \bibnamefont
  {Rodrigues}}, \bibinfo {author} {\bibfnamefont {N.}~\bibnamefont {Sousa}},
  \emph {et~al.},\ }\href@noop {} {\bibfield  {journal} {\bibinfo  {journal}
  {Physical Review Letters}\ }\textbf {\bibinfo {volume} {122}},\ \bibinfo
  {pages} {208101} (\bibinfo {year} {2019})}\BibitemShut {NoStop}%
\bibitem [{\citenamefont {Kartvelishvili}\ \emph {et~al.}(2021)\citenamefont
  {Kartvelishvili}, \citenamefont {Khoury},\ and\ \citenamefont
  {Sharma}}]{kartvelishvili2021self}%
  \BibitemOpen
  \bibfield  {author} {\bibinfo {author} {\bibfnamefont {G.}~\bibnamefont
  {Kartvelishvili}}, \bibinfo {author} {\bibfnamefont {J.}~\bibnamefont
  {Khoury}},\ and\ \bibinfo {author} {\bibfnamefont {A.}~\bibnamefont
  {Sharma}},\ }\href@noop {} {\bibfield  {journal} {\bibinfo  {journal}
  {Journal of Cosmology and Astroparticle Physics}\ }\textbf {\bibinfo {volume}
  {2021}}\bibinfo  {number} { (02)},\ \bibinfo {pages} {028}}\BibitemShut
  {NoStop}%
\bibitem [{\citenamefont {Bak}\ \emph {et~al.}(1987)\citenamefont {Bak},
  \citenamefont {Tang},\ and\ \citenamefont {Wiesenfeld}}]{BTW1987}%
  \BibitemOpen
\bibfield  {number} {  }\bibfield  {author} {\bibinfo {author} {\bibfnamefont
  {P.}~\bibnamefont {Bak}}, \bibinfo {author} {\bibfnamefont {C.}~\bibnamefont
  {Tang}},\ and\ \bibinfo {author} {\bibfnamefont {K.}~\bibnamefont
  {Wiesenfeld}},\ }\href@noop {} {\bibfield  {journal} {\bibinfo  {journal}
  {Physical Review Letters}\ }\textbf {\bibinfo {volume} {59}},\ \bibinfo
  {pages} {381} (\bibinfo {year} {1987})}\BibitemShut {NoStop}%
\bibitem [{\citenamefont {Dickman}\ \emph {et~al.}(1998)\citenamefont
  {Dickman}, \citenamefont {Vespignani},\ and\ \citenamefont
  {Zapperi}}]{dickman1998absorbingSOC}%
  \BibitemOpen
  \bibfield  {author} {\bibinfo {author} {\bibfnamefont {R.}~\bibnamefont
  {Dickman}}, \bibinfo {author} {\bibfnamefont {A.}~\bibnamefont
  {Vespignani}},\ and\ \bibinfo {author} {\bibfnamefont {S.}~\bibnamefont
  {Zapperi}},\ }\href@noop {} {\bibfield  {journal} {\bibinfo  {journal}
  {Physical Review E}\ }\textbf {\bibinfo {volume} {57}},\ \bibinfo {pages}
  {5095} (\bibinfo {year} {1998})}\BibitemShut {NoStop}%
\bibitem [{\citenamefont {Pruessner}(2012)}]{pruessner2012self}%
  \BibitemOpen
  \bibfield  {author} {\bibinfo {author} {\bibfnamefont {G.}~\bibnamefont
  {Pruessner}},\ }\href@noop {} {\emph {\bibinfo {title} {Self-organised
  criticality: theory, models and characterisation}}}\ (\bibinfo  {publisher}
  {Cambridge University Press},\ \bibinfo {year} {2012})\BibitemShut {NoStop}%
\bibitem [{\citenamefont {Zapperi}(2022)}]{zapperi2022crackling}%
  \BibitemOpen
  \bibfield  {author} {\bibinfo {author} {\bibfnamefont {S.}~\bibnamefont
  {Zapperi}},\ }\href@noop {} {\emph {\bibinfo {title} {Crackling Noise:
  Statistical Physics of Avalanche Phenomena}}}\ (\bibinfo  {publisher} {Oxford
  University Press},\ \bibinfo {year} {2022})\BibitemShut {NoStop}%
\bibitem [{\citenamefont {Sornette}(2009)}]{sornette2009dragon_kings}%
  \BibitemOpen
  \bibfield  {author} {\bibinfo {author} {\bibfnamefont {D.}~\bibnamefont
  {Sornette}},\ }\href@noop {} {\bibfield  {journal} {\bibinfo  {journal}
  {International Journal of Terraspace Science and Engineering}\ }\textbf
  {\bibinfo {volume} {2}},\ \bibinfo {pages} {1} (\bibinfo {year}
  {2009})}\BibitemShut {NoStop}%
\bibitem [{\citenamefont {Sornette}\ and\ \citenamefont
  {Ouillon}(2012)}]{sornette2012dragon}%
  \BibitemOpen
  \bibfield  {author} {\bibinfo {author} {\bibfnamefont {D.}~\bibnamefont
  {Sornette}}\ and\ \bibinfo {author} {\bibfnamefont {G.}~\bibnamefont
  {Ouillon}},\ }\href@noop {} {\bibfield  {journal} {\bibinfo  {journal} {The
  European Physical Journal Special Topics}\ }\textbf {\bibinfo {volume}
  {205}},\ \bibinfo {pages} {1} (\bibinfo {year} {2012})}\BibitemShut {NoStop}%
\bibitem [{\citenamefont {Filimonov}\ and\ \citenamefont
  {Sornette}(2015)}]{sornette2015financial_dk}%
  \BibitemOpen
  \bibfield  {author} {\bibinfo {author} {\bibfnamefont {V.}~\bibnamefont
  {Filimonov}}\ and\ \bibinfo {author} {\bibfnamefont {D.}~\bibnamefont
  {Sornette}},\ }\href@noop {} {\bibfield  {journal} {\bibinfo  {journal}
  {Chaos, Solitons \& Fractals}\ }\textbf {\bibinfo {volume} {74}},\ \bibinfo
  {pages} {27} (\bibinfo {year} {2015})}\BibitemShut {NoStop}%
\bibitem [{\citenamefont {Wheatley}\ \emph {et~al.}(2017)\citenamefont
  {Wheatley}, \citenamefont {Sovacool},\ and\ \citenamefont
  {Sornette}}]{sornette2017nuclear_dk}%
  \BibitemOpen
  \bibfield  {author} {\bibinfo {author} {\bibfnamefont {S.}~\bibnamefont
  {Wheatley}}, \bibinfo {author} {\bibfnamefont {B.}~\bibnamefont {Sovacool}},\
  and\ \bibinfo {author} {\bibfnamefont {D.}~\bibnamefont {Sornette}},\
  }\href@noop {} {\bibfield  {journal} {\bibinfo  {journal} {Risk Analysis}\
  }\textbf {\bibinfo {volume} {37}},\ \bibinfo {pages} {99} (\bibinfo {year}
  {2017})}\BibitemShut {NoStop}%
\bibitem [{\citenamefont {Pisarenko}\ and\ \citenamefont
  {Sornette}(2012)}]{sornette2012city_dk}%
  \BibitemOpen
  \bibfield  {author} {\bibinfo {author} {\bibfnamefont {V.~F.}\ \bibnamefont
  {Pisarenko}}\ and\ \bibinfo {author} {\bibfnamefont {D.}~\bibnamefont
  {Sornette}},\ }\href@noop {} {\bibfield  {journal} {\bibinfo  {journal} {The
  European Physical Journal Special Topics}\ }\textbf {\bibinfo {volume}
  {205}},\ \bibinfo {pages} {95} (\bibinfo {year} {2012})}\BibitemShut
  {NoStop}%
\bibitem [{\citenamefont {Bochdansky}\ \emph {et~al.}(2016)\citenamefont
  {Bochdansky}, \citenamefont {Clouse},\ and\ \citenamefont
  {Herndl}}]{bochdansky2016deep_sea}%
  \BibitemOpen
  \bibfield  {author} {\bibinfo {author} {\bibfnamefont {A.~B.}\ \bibnamefont
  {Bochdansky}}, \bibinfo {author} {\bibfnamefont {M.~A.}\ \bibnamefont
  {Clouse}},\ and\ \bibinfo {author} {\bibfnamefont {G.~J.}\ \bibnamefont
  {Herndl}},\ }\href@noop {} {\bibfield  {journal} {\bibinfo  {journal}
  {Scientific Reports}\ }\textbf {\bibinfo {volume} {6}},\ \bibinfo {pages} {1}
  (\bibinfo {year} {2016})}\BibitemShut {NoStop}%
\bibitem [{\citenamefont {De~Arcangelis}(2012)}]{de2012dungeons}%
  \BibitemOpen
  \bibfield  {author} {\bibinfo {author} {\bibfnamefont {L.}~\bibnamefont
  {De~Arcangelis}},\ }\href@noop {} {\bibfield  {journal} {\bibinfo  {journal}
  {The European Physical Journal Special Topics}\ }\textbf {\bibinfo {volume}
  {205}},\ \bibinfo {pages} {243} (\bibinfo {year} {2012})}\BibitemShut
  {NoStop}%
\bibitem [{\citenamefont {Mishra}\ \emph {et~al.}(2018)\citenamefont {Mishra},
  \citenamefont {Saha}, \citenamefont {Vigneshwaran}, \citenamefont {Pal},
  \citenamefont {Kapitaniak},\ and\ \citenamefont {Dana}}]{mishra2018dragon}%
  \BibitemOpen
  \bibfield  {author} {\bibinfo {author} {\bibfnamefont {A.}~\bibnamefont
  {Mishra}}, \bibinfo {author} {\bibfnamefont {S.}~\bibnamefont {Saha}},
  \bibinfo {author} {\bibfnamefont {M.}~\bibnamefont {Vigneshwaran}}, \bibinfo
  {author} {\bibfnamefont {P.}~\bibnamefont {Pal}}, \bibinfo {author}
  {\bibfnamefont {T.}~\bibnamefont {Kapitaniak}},\ and\ \bibinfo {author}
  {\bibfnamefont {S.~K.}\ \bibnamefont {Dana}},\ }\href@noop {} {\bibfield
  {journal} {\bibinfo  {journal} {Physical Review E}\ }\textbf {\bibinfo
  {volume} {97}},\ \bibinfo {pages} {062311} (\bibinfo {year}
  {2018})}\BibitemShut {NoStop}%
\bibitem [{\citenamefont {Premraj}\ \emph {et~al.}(2021)\citenamefont
  {Premraj}, \citenamefont {Suresh}, \citenamefont {Pawar}, \citenamefont
  {Kabiraj}, \citenamefont {Prasad},\ and\ \citenamefont
  {Sujith}}]{premraj2021catastrophic_transition}%
  \BibitemOpen
  \bibfield  {author} {\bibinfo {author} {\bibfnamefont {D.}~\bibnamefont
  {Premraj}}, \bibinfo {author} {\bibfnamefont {K.}~\bibnamefont {Suresh}},
  \bibinfo {author} {\bibfnamefont {S.~A.}\ \bibnamefont {Pawar}}, \bibinfo
  {author} {\bibfnamefont {L.}~\bibnamefont {Kabiraj}}, \bibinfo {author}
  {\bibfnamefont {A.}~\bibnamefont {Prasad}},\ and\ \bibinfo {author}
  {\bibfnamefont {R.}~\bibnamefont {Sujith}},\ }\href@noop {} {\bibfield
  {journal} {\bibinfo  {journal} {EPL (Europhysics Letters)}\ }\textbf
  {\bibinfo {volume} {134}},\ \bibinfo {pages} {34006} (\bibinfo {year}
  {2021})}\BibitemShut {NoStop}%
\bibitem [{\citenamefont {Sachs}\ \emph {et~al.}(2012)\citenamefont {Sachs},
  \citenamefont {Yoder}, \citenamefont {Turcotte}, \citenamefont {Rundle},\
  and\ \citenamefont {Malamud}}]{sachs2012examples_and_soc}%
  \BibitemOpen
  \bibfield  {author} {\bibinfo {author} {\bibfnamefont {M.~K.}\ \bibnamefont
  {Sachs}}, \bibinfo {author} {\bibfnamefont {M.~R.}\ \bibnamefont {Yoder}},
  \bibinfo {author} {\bibfnamefont {D.~L.}\ \bibnamefont {Turcotte}}, \bibinfo
  {author} {\bibfnamefont {J.~B.}\ \bibnamefont {Rundle}},\ and\ \bibinfo
  {author} {\bibfnamefont {B.~D.}\ \bibnamefont {Malamud}},\ }\href@noop {}
  {\bibfield  {journal} {\bibinfo  {journal} {The European Physical Journal
  Special Topics}\ }\textbf {\bibinfo {volume} {205}},\ \bibinfo {pages} {167}
  (\bibinfo {year} {2012})}\BibitemShut {NoStop}%
\bibitem [{\citenamefont {Grassberger}(1993)}]{grassberger1993forest_fire}%
  \BibitemOpen
  \bibfield  {author} {\bibinfo {author} {\bibfnamefont {P.}~\bibnamefont
  {Grassberger}},\ }\href@noop {} {\bibfield  {journal} {\bibinfo  {journal}
  {Journal of Physics A: Mathematical and General}\ }\textbf {\bibinfo {volume}
  {26}},\ \bibinfo {pages} {2081} (\bibinfo {year} {1993})}\BibitemShut
  {NoStop}%
\bibitem [{\citenamefont {Zapperi}\ \emph {et~al.}(1995)\citenamefont
  {Zapperi}, \citenamefont {Lauritsen},\ and\ \citenamefont
  {Stanley}}]{zapperi1995sobp}%
  \BibitemOpen
  \bibfield  {author} {\bibinfo {author} {\bibfnamefont {S.}~\bibnamefont
  {Zapperi}}, \bibinfo {author} {\bibfnamefont {K.~B.}\ \bibnamefont
  {Lauritsen}},\ and\ \bibinfo {author} {\bibfnamefont {H.~E.}\ \bibnamefont
  {Stanley}},\ }\href@noop {} {\bibfield  {journal} {\bibinfo  {journal}
  {Physical Review Letters}\ }\textbf {\bibinfo {volume} {75}},\ \bibinfo
  {pages} {4071} (\bibinfo {year} {1995})}\BibitemShut {NoStop}%
\bibitem [{\citenamefont {Amaral}\ and\ \citenamefont
  {Lauritsen}(1996)}]{amaral1996self}%
  \BibitemOpen
  \bibfield  {author} {\bibinfo {author} {\bibfnamefont {L.~A.~N.}\
  \bibnamefont {Amaral}}\ and\ \bibinfo {author} {\bibfnamefont {K.~B.}\
  \bibnamefont {Lauritsen}},\ }\href@noop {} {\bibfield  {journal} {\bibinfo
  {journal} {Physical Review E}\ }\textbf {\bibinfo {volume} {54}},\ \bibinfo
  {pages} {R4512} (\bibinfo {year} {1996})}\BibitemShut {NoStop}%
\bibitem [{\citenamefont {Watanabe}\ \emph {et~al.}(2015)\citenamefont
  {Watanabe}, \citenamefont {Mizutaka},\ and\ \citenamefont
  {Yakubo}}]{watanabe2015fractal}%
  \BibitemOpen
  \bibfield  {author} {\bibinfo {author} {\bibfnamefont {A.}~\bibnamefont
  {Watanabe}}, \bibinfo {author} {\bibfnamefont {S.}~\bibnamefont {Mizutaka}},\
  and\ \bibinfo {author} {\bibfnamefont {K.}~\bibnamefont {Yakubo}},\
  }\href@noop {} {\bibfield  {journal} {\bibinfo  {journal} {Journal of the
  Physical Society of Japan}\ }\textbf {\bibinfo {volume} {84}},\ \bibinfo
  {pages} {114003} (\bibinfo {year} {2015})}\BibitemShut {NoStop}%
\bibitem [{\citenamefont {Lin}\ \emph {et~al.}(2018)\citenamefont {Lin},
  \citenamefont {Burghardt}, \citenamefont {Rohden}, \citenamefont {No{\"e}l},\
  and\ \citenamefont {D'Souza}}]{dsouza2018sodk}%
  \BibitemOpen
  \bibfield  {author} {\bibinfo {author} {\bibfnamefont {Y.}~\bibnamefont
  {Lin}}, \bibinfo {author} {\bibfnamefont {K.}~\bibnamefont {Burghardt}},
  \bibinfo {author} {\bibfnamefont {M.}~\bibnamefont {Rohden}}, \bibinfo
  {author} {\bibfnamefont {P.-A.}\ \bibnamefont {No{\"e}l}},\ and\ \bibinfo
  {author} {\bibfnamefont {R.~M.}\ \bibnamefont {D'Souza}},\ }\href@noop {}
  {\bibfield  {journal} {\bibinfo  {journal} {Physical Review E}\ }\textbf
  {\bibinfo {volume} {98}},\ \bibinfo {pages} {022127} (\bibinfo {year}
  {2018})}\BibitemShut {NoStop}%
\bibitem [{\citenamefont {Kinouchi}\ \emph {et~al.}(2019)\citenamefont
  {Kinouchi}, \citenamefont {Brochini}, \citenamefont {Costa}, \citenamefont
  {Campos},\ and\ \citenamefont {Copelli}}]{kinouchi2019stochastic}%
  \BibitemOpen
  \bibfield  {author} {\bibinfo {author} {\bibfnamefont {O.}~\bibnamefont
  {Kinouchi}}, \bibinfo {author} {\bibfnamefont {L.}~\bibnamefont {Brochini}},
  \bibinfo {author} {\bibfnamefont {A.~A.}\ \bibnamefont {Costa}}, \bibinfo
  {author} {\bibfnamefont {J.~G.~F.}\ \bibnamefont {Campos}},\ and\ \bibinfo
  {author} {\bibfnamefont {M.}~\bibnamefont {Copelli}},\ }\href@noop {}
  {\bibfield  {journal} {\bibinfo  {journal} {Scientific reports}\ }\textbf
  {\bibinfo {volume} {9}},\ \bibinfo {pages} {1} (\bibinfo {year}
  {2019})}\BibitemShut {NoStop}%
\bibitem [{\citenamefont {Cavalcante}\ \emph {et~al.}(2013)\citenamefont
  {Cavalcante}, \citenamefont {Ori{\'a}}, \citenamefont {Sornette},
  \citenamefont {Ott},\ and\ \citenamefont
  {Gauthier}}]{cavalcante2013predictability}%
  \BibitemOpen
  \bibfield  {author} {\bibinfo {author} {\bibfnamefont {H.~L. d.~S.}\
  \bibnamefont {Cavalcante}}, \bibinfo {author} {\bibfnamefont
  {M.}~\bibnamefont {Ori{\'a}}}, \bibinfo {author} {\bibfnamefont
  {D.}~\bibnamefont {Sornette}}, \bibinfo {author} {\bibfnamefont
  {E.}~\bibnamefont {Ott}},\ and\ \bibinfo {author} {\bibfnamefont {D.~J.}\
  \bibnamefont {Gauthier}},\ }\href@noop {} {\bibfield  {journal} {\bibinfo
  {journal} {Physical Review Letters}\ }\textbf {\bibinfo {volume} {111}},\
  \bibinfo {pages} {198701} (\bibinfo {year} {2013})}\BibitemShut {NoStop}%
\bibitem [{\citenamefont {De~Queiroz}\ and\ \citenamefont
  {Bahiana}(2001)}]{queiroz2001barkhausen}%
  \BibitemOpen
  \bibfield  {author} {\bibinfo {author} {\bibfnamefont {S.}~\bibnamefont
  {De~Queiroz}}\ and\ \bibinfo {author} {\bibfnamefont {M.}~\bibnamefont
  {Bahiana}},\ }\href@noop {} {\bibfield  {journal} {\bibinfo  {journal}
  {Physical Review E}\ }\textbf {\bibinfo {volume} {64}},\ \bibinfo {pages}
  {066127} (\bibinfo {year} {2001})}\BibitemShut {NoStop}%
\bibitem [{\citenamefont {Hwa}\ and\ \citenamefont
  {Kardar}(1992)}]{hwa1992running}%
  \BibitemOpen
  \bibfield  {author} {\bibinfo {author} {\bibfnamefont {T.}~\bibnamefont
  {Hwa}}\ and\ \bibinfo {author} {\bibfnamefont {M.}~\bibnamefont {Kardar}},\
  }\href@noop {} {\bibfield  {journal} {\bibinfo  {journal} {Physical Review
  A}\ }\textbf {\bibinfo {volume} {45}},\ \bibinfo {pages} {7002} (\bibinfo
  {year} {1992})}\BibitemShut {NoStop}%
\bibitem [{\citenamefont {Pradhan}(2021)}]{pradhan2021time}%
  \BibitemOpen
  \bibfield  {author} {\bibinfo {author} {\bibfnamefont {P.}~\bibnamefont
  {Pradhan}},\ }\href@noop {} {\bibfield  {journal} {\bibinfo  {journal}
  {Frontiers in Physics}\ }\textbf {\bibinfo {volume} {9}},\ \bibinfo {pages}
  {641233} (\bibinfo {year} {2021})}\BibitemShut {NoStop}%
\bibitem [{\citenamefont {Vespignani}\ and\ \citenamefont
  {Zapperi}(1998)}]{vespignani1998soc_mft}%
  \BibitemOpen
  \bibfield  {author} {\bibinfo {author} {\bibfnamefont {A.}~\bibnamefont
  {Vespignani}}\ and\ \bibinfo {author} {\bibfnamefont {S.}~\bibnamefont
  {Zapperi}},\ }\href@noop {} {\bibfield  {journal} {\bibinfo  {journal}
  {Physical Review E}\ }\textbf {\bibinfo {volume} {57}},\ \bibinfo {pages}
  {6345} (\bibinfo {year} {1998})}\BibitemShut {NoStop}%
\bibitem [{\citenamefont {Dickman}\ \emph {et~al.}(2000)\citenamefont
  {Dickman}, \citenamefont {Mu{\~n}oz}, \citenamefont {Vespignani},\ and\
  \citenamefont {Zapperi}}]{dickman2000paths}%
  \BibitemOpen
  \bibfield  {author} {\bibinfo {author} {\bibfnamefont {R.}~\bibnamefont
  {Dickman}}, \bibinfo {author} {\bibfnamefont {M.~A.}\ \bibnamefont
  {Mu{\~n}oz}}, \bibinfo {author} {\bibfnamefont {A.}~\bibnamefont
  {Vespignani}},\ and\ \bibinfo {author} {\bibfnamefont {S.}~\bibnamefont
  {Zapperi}},\ }\href@noop {} {\bibfield  {journal} {\bibinfo  {journal}
  {Brazilian Journal of Physics}\ }\textbf {\bibinfo {volume} {30}},\ \bibinfo
  {pages} {27} (\bibinfo {year} {2000})}\BibitemShut {NoStop}%
\bibitem [{\citenamefont {Cafiero}\ \emph {et~al.}(1995)\citenamefont
  {Cafiero}, \citenamefont {Loreto}, \citenamefont {Pietronero}, \citenamefont
  {Vespignani},\ and\ \citenamefont {Zapperi}}]{cafiero1995local}%
  \BibitemOpen
  \bibfield  {author} {\bibinfo {author} {\bibfnamefont {R.}~\bibnamefont
  {Cafiero}}, \bibinfo {author} {\bibfnamefont {V.}~\bibnamefont {Loreto}},
  \bibinfo {author} {\bibfnamefont {L.}~\bibnamefont {Pietronero}}, \bibinfo
  {author} {\bibfnamefont {A.}~\bibnamefont {Vespignani}},\ and\ \bibinfo
  {author} {\bibfnamefont {S.}~\bibnamefont {Zapperi}},\ }\href@noop {}
  {\bibfield  {journal} {\bibinfo  {journal} {EPL (Europhysics Letters)}\
  }\textbf {\bibinfo {volume} {29}},\ \bibinfo {pages} {111} (\bibinfo {year}
  {1995})}\BibitemShut {NoStop}%
\bibitem [{\citenamefont {Bonachela}\ and\ \citenamefont
  {Mu{\~n}oz}(2009)}]{munoz2009nonconservative}%
  \BibitemOpen
  \bibfield  {author} {\bibinfo {author} {\bibfnamefont {J.~A.}\ \bibnamefont
  {Bonachela}}\ and\ \bibinfo {author} {\bibfnamefont {M.~A.}\ \bibnamefont
  {Mu{\~n}oz}},\ }\href@noop {} {\bibfield  {journal} {\bibinfo  {journal}
  {Journal of Statistical Mechanics: Theory and Experiment}\ }\textbf {\bibinfo
  {volume} {2009}},\ \bibinfo {pages} {P09009} (\bibinfo {year}
  {2009})}\BibitemShut {NoStop}%
\bibitem [{\citenamefont {No{\"e}l}\ \emph {et~al.}(2013)\citenamefont
  {No{\"e}l}, \citenamefont {Brummitt},\ and\ \citenamefont
  {D’Souza}}]{noel2013controlling}%
  \BibitemOpen
  \bibfield  {author} {\bibinfo {author} {\bibfnamefont {P.-A.}\ \bibnamefont
  {No{\"e}l}}, \bibinfo {author} {\bibfnamefont {C.~D.}\ \bibnamefont
  {Brummitt}},\ and\ \bibinfo {author} {\bibfnamefont {R.~M.}\ \bibnamefont
  {D’Souza}},\ }\href@noop {} {\bibfield  {journal} {\bibinfo  {journal}
  {Physical Review Letters}\ }\textbf {\bibinfo {volume} {111}},\ \bibinfo
  {pages} {078701} (\bibinfo {year} {2013})}\BibitemShut {NoStop}%
\bibitem [{\citenamefont {di~Santo}\ \emph {et~al.}(2016)\citenamefont
  {di~Santo}, \citenamefont {Burioni}, \citenamefont {Vezzani},\ and\
  \citenamefont {Mu{\~n}oz}}]{munoz2016so_bistability}%
  \BibitemOpen
  \bibfield  {author} {\bibinfo {author} {\bibfnamefont {S.}~\bibnamefont
  {di~Santo}}, \bibinfo {author} {\bibfnamefont {R.}~\bibnamefont {Burioni}},
  \bibinfo {author} {\bibfnamefont {A.}~\bibnamefont {Vezzani}},\ and\ \bibinfo
  {author} {\bibfnamefont {M.~A.}\ \bibnamefont {Mu{\~n}oz}},\ }\href@noop {}
  {\bibfield  {journal} {\bibinfo  {journal} {Physical Review Letters}\
  }\textbf {\bibinfo {volume} {116}},\ \bibinfo {pages} {240601} (\bibinfo
  {year} {2016})}\BibitemShut {NoStop}%
\bibitem [{\citenamefont {Buend{\'\i}a}\ \emph {et~al.}(2020)\citenamefont
  {Buend{\'\i}a}, \citenamefont {di~Santo}, \citenamefont {Villegas},
  \citenamefont {Burioni},\ and\ \citenamefont
  {Mu{\~n}oz}}]{munoz2020sob_brain}%
  \BibitemOpen
  \bibfield  {author} {\bibinfo {author} {\bibfnamefont {V.}~\bibnamefont
  {Buend{\'\i}a}}, \bibinfo {author} {\bibfnamefont {S.}~\bibnamefont
  {di~Santo}}, \bibinfo {author} {\bibfnamefont {P.}~\bibnamefont {Villegas}},
  \bibinfo {author} {\bibfnamefont {R.}~\bibnamefont {Burioni}},\ and\ \bibinfo
  {author} {\bibfnamefont {M.~A.}\ \bibnamefont {Mu{\~n}oz}},\ }\href@noop {}
  {\bibfield  {journal} {\bibinfo  {journal} {Physical Review Research}\
  }\textbf {\bibinfo {volume} {2}},\ \bibinfo {pages} {013318} (\bibinfo {year}
  {2020})}\BibitemShut {NoStop}%
\bibitem [{\citenamefont {Gil}\ and\ \citenamefont
  {Sornette}(1996)}]{sornette1996landau-ginsburg-soc}%
  \BibitemOpen
  \bibfield  {author} {\bibinfo {author} {\bibfnamefont {L.}~\bibnamefont
  {Gil}}\ and\ \bibinfo {author} {\bibfnamefont {D.}~\bibnamefont {Sornette}},\
  }\href@noop {} {\bibfield  {journal} {\bibinfo  {journal} {Physical Review
  Letters}\ }\textbf {\bibinfo {volume} {76}},\ \bibinfo {pages} {3991}
  (\bibinfo {year} {1996})}\BibitemShut {NoStop}%
\bibitem [{\citenamefont {Mikaberidze}\ and\ \citenamefont
  {D’Souza}(2022)}]{mikaberidze2022sandpile}%
  \BibitemOpen
  \bibfield  {author} {\bibinfo {author} {\bibfnamefont {G.}~\bibnamefont
  {Mikaberidze}}\ and\ \bibinfo {author} {\bibfnamefont {R.~M.}\ \bibnamefont
  {D’Souza}},\ }\href@noop {} {\bibfield  {journal} {\bibinfo  {journal}
  {Chaos: An Interdisciplinary Journal of Nonlinear Science}\ }\textbf
  {\bibinfo {volume} {32}},\ \bibinfo {pages} {053121} (\bibinfo {year}
  {2022})}\BibitemShut {NoStop}%
\bibitem [{\citenamefont {White}\ and\ \citenamefont
  {Dahmen}(2003)}]{white2003driving}%
  \BibitemOpen
  \bibfield  {author} {\bibinfo {author} {\bibfnamefont {R.~A.}\ \bibnamefont
  {White}}\ and\ \bibinfo {author} {\bibfnamefont {K.~A.}\ \bibnamefont
  {Dahmen}},\ }\href@noop {} {\bibfield  {journal} {\bibinfo  {journal}
  {Physical Review Letters}\ }\textbf {\bibinfo {volume} {91}},\ \bibinfo
  {pages} {085702} (\bibinfo {year} {2003})}\BibitemShut {NoStop}%
\bibitem [{\citenamefont {Pathria}(2016)}]{pathria2016statistical}%
  \BibitemOpen
  \bibfield  {author} {\bibinfo {author} {\bibfnamefont {R.~K.}\ \bibnamefont
  {Pathria}},\ }\href@noop {} {\emph {\bibinfo {title} {Statistical
  mechanics}}}\ (\bibinfo  {publisher} {Elsevier},\ \bibinfo {year}
  {2016})\BibitemShut {NoStop}%
\bibitem [{Note1()}]{Note1}%
  \BibitemOpen
  \bibinfo {note} {Note that driving can be independent of perturbation. In the
  self-organized branching process, \cite {zapperi1995sobp} the driving
  increments the branching probability while perturbation is introduced by
  activating a single site.}\BibitemShut {Stop}%
\bibitem [{\citenamefont {Eliazar}(2017)}]{eliazar2017black}%
  \BibitemOpen
  \bibfield  {author} {\bibinfo {author} {\bibfnamefont {I.}~\bibnamefont
  {Eliazar}},\ }\href@noop {} {\bibfield  {journal} {\bibinfo  {journal} {EPL
  (Europhysics Letters)}\ }\textbf {\bibinfo {volume} {119}},\ \bibinfo {pages}
  {60007} (\bibinfo {year} {2017})}\BibitemShut {NoStop}%
\bibitem [{\citenamefont {Janczura}\ and\ \citenamefont
  {Weron}(2012)}]{janczura2012black}%
  \BibitemOpen
  \bibfield  {author} {\bibinfo {author} {\bibfnamefont {J.}~\bibnamefont
  {Janczura}}\ and\ \bibinfo {author} {\bibfnamefont {R.}~\bibnamefont
  {Weron}},\ }\href@noop {} {\bibfield  {journal} {\bibinfo  {journal} {The
  European Physical Journal Special Topics}\ }\textbf {\bibinfo {volume}
  {205}},\ \bibinfo {pages} {79} (\bibinfo {year} {2012})}\BibitemShut
  {NoStop}%
\bibitem [{\citenamefont {Cardy}\ and\ \citenamefont
  {Grassberger}(1985)}]{cardy1985epidemic}%
  \BibitemOpen
  \bibfield  {author} {\bibinfo {author} {\bibfnamefont {J.~L.}\ \bibnamefont
  {Cardy}}\ and\ \bibinfo {author} {\bibfnamefont {P.}~\bibnamefont
  {Grassberger}},\ }\href@noop {} {\bibfield  {journal} {\bibinfo  {journal}
  {Journal of Physics A: Mathematical and General}\ }\textbf {\bibinfo {volume}
  {18}},\ \bibinfo {pages} {L267} (\bibinfo {year} {1985})}\BibitemShut
  {NoStop}%
\bibitem [{Note2()}]{Note2}%
  \BibitemOpen
  \bibinfo {note} {Notice that the driving impulse is $\Delta E=N^{-1}$ for
  both models of \cite {dsouza2018sodk}. Thus the second-order DKs disappear in
  the thermodynamic limit unless $\epsilon $ also scales with $N$ (inoculation
  model). Yet, for the first-order DKs (complex contagion model), the peak
  remains sharply pronounced in the thermodynamic limit since it does not
  require the system to become strongly supercritical.}\BibitemShut {Stop}%
\bibitem [{\citenamefont {Ódor}\ \emph {et~al.}(2022)\citenamefont {Ódor},
  \citenamefont {Deng}, \citenamefont {Hartmann},\ and\ \citenamefont
  {Kelling}}]{odor2022powergrid}%
  \BibitemOpen
  \bibfield  {author} {\bibinfo {author} {\bibfnamefont {G.}~\bibnamefont
  {Ódor}}, \bibinfo {author} {\bibfnamefont {S.}~\bibnamefont {Deng}},
  \bibinfo {author} {\bibfnamefont {B.}~\bibnamefont {Hartmann}},\ and\
  \bibinfo {author} {\bibfnamefont {J.}~\bibnamefont {Kelling}},\ }\href
  {https://doi.org/10.48550/ARXIV.2205.13472} {\bibinfo {title}
  {Synchronization dynamics on the {EU} and {US }power grids}} (\bibinfo {year}
  {2022})\BibitemShut {NoStop}%
\end{thebibliography}%

\end{document}


\title{Supplemental Information}
	\author{}
	\date{}
	\maketitle
	
	The purpose of this document is to show the details of derivations performed for the manuscript under review. 
	
	\section{Deriving Eq. (4) in the manuscript}
	We start with the expression for $S_{DK}$ (Eq. (3) in the manuscript).
	
	\begin{equation}\label{sdk}
			S_{DK} = \sum_{j=0}^\infty     \frac{j\Delta E}{\epsilon}    (1-p(E_c+j\Delta E))\prod_{i=0}^{j-1} p(E_c+i\Delta E).
	\end{equation}
	Here $p(E)$ is given by Eq. (1) in the manuscript
	\begin{equation}\label{p(E)}
		p(E) = \begin{cases}
			1											& \text{for } E\le E_c \\
			1-k(E-E_c)^\lambda 		& \text{for } E>E_c.
		\end{cases}
	\end{equation}
	Making the substitution of Eq. \eqref{p(E)} into Eq. \eqref{sdk} we get
	\begin{equation}\label{sdk2}
		\begin{split}
		S_{DK} &= \sum_{j=0}^\infty     \frac{j\Delta E}{\epsilon}   k(j\Delta E)^\lambda 	\prod_{i=0}^{j-1} (1-k(i\Delta E)^\lambda ) \\
		&= \sum_{j=0}^\infty     \frac{k}{\epsilon}   (j\Delta E)^{\lambda+1} 	\prod_{i=0}^{j-1} (1-k(i\Delta E)^\lambda )
		\end{split}
	\end{equation}
	Let us compute the product term first:
	\begin{equation}
		\begin{split}
		\prod_{i=0}^{j-1} (1-k(i\Delta E)^\lambda ) = \exp\left(\sum_{i=0}^{j-1} \ln(1-k(i\Delta E)^\lambda )\right)
		\approx
		\exp\left(-\sum_{i=0}^{j-1} k(i\Delta E)^\lambda )\right)
		\\
		\approx
		\exp\left(-k\Delta E^\lambda\int_0^{j-1} i^\lambda di\right)
		=
		\exp\left(-\frac{k\Delta E^\lambda}{\lambda+1}(j-1)^{\lambda+1}\right)
	\end{split}
	\end{equation}
	Here in the second equality we used taylor expansion in $\Delta E$. Note that this expansion breaks down for very large values $i$, but these terms do not contribute significantly since they correspond to very large and unlikely excursions into the supercritical phase. In the third equality we used an approximation of the sum with an integral. For convenience, we define $A=\frac{k\Delta E^\lambda}{\lambda+1}$, and substitute our result into Eq. \eqref{sdk2}
		\begin{equation}\label{sdk3}
		\begin{split}
			S_{DK} \approx 
			\sum_{j=0}^\infty     \frac{k}{\epsilon}   (j\Delta E)^{\lambda+1} e^{-A(j-1)^{\lambda+1}}
			\approx
			\int_{0}^\infty     \frac{k}{\epsilon}   (j\Delta E)^{\lambda+1} e^{-A(j-1)^{\lambda+1}}dj \\
			=
			\frac{k}{\epsilon} \Delta E^{\lambda+1} \int_{0}^\infty j^{\lambda+1} e^{-A(j-1)^{\lambda+1}}dj
		\end{split}
	\end{equation}
	Since $A\propto \Delta E^\lambda<<1$, the exponent will be significant only for large values of $j$, and we can neglect the $-1$ compared to this large $j$.
	\begin{equation}\label{sdk4}
		\begin{split}
			S_{DK} &\approx 
			\frac{k}{\epsilon} \Delta E^{\lambda+1} \int_{0}^\infty j^{\lambda+1} e^{-Aj^{\lambda+1}}dj \\
			&= 
			\frac{k}{\epsilon} \Delta E^{\lambda+1}
			\frac{1}{\lambda+1}A^{-\frac{\lambda+2}{\lambda+1}} \int_{0}^\infty z^\frac{1}{\lambda+1} e^{-z}dz \\
			&=
			\frac{k}{\epsilon} \Delta E^{\lambda+1}
			\frac{1}{\lambda+1}\left(\frac{k\Delta E^\lambda}{\lambda+1}\right)^{-\frac{\lambda+2}{\lambda+1}} \Gamma\left(\frac{1}{\lambda+1}+1\right) \\
			&=
			\frac{1}{\epsilon}\left(\frac{\lambda+1}{k}\Delta E\right)^{\frac{1}{1+\lambda}} \Gamma\left(\frac{\lambda+2}{\lambda+1}\right)
		\end{split}
	\end{equation}
	In the second line we changed the integration variable to $z=Aj^{\lambda+1}$. The validity of this approximation can be easily verified by comparing its predictions with numerical evaluation of the original expression.

	\section{Deriving Eq. (7) in the manuscript}
	To obtain Eq. (7) in the manuscript, we start with the requirement that the two scales $S_{DK}$ and $S_\epsilon$ scale identically $S_{DK}\sim S_\epsilon$ in the limit $\epsilon\rightarrow 0$ and $\Delta E\rightarrow 0$. The two scales are given by
	\begin{equation}\label{s_DK}
		S_{DK} \approx \frac{1}{\epsilon}\left( \frac{1+\lambda}{k}\Delta E\right)^{\frac{1}{1+\lambda}}\Gamma\left( \frac{2+\lambda}{1+\lambda} \right)
	\end{equation}
	\begin{equation}\label{s_eps}
		S_\epsilon \approx \left(\int_0^\infty 	z^{1-\tau}G(z)dz\right)^{-\frac{1}{2-\tau}} \left(\frac{\Delta E}{\epsilon}\right)^\frac{1}{2-\tau}.
	\end{equation}
	Neglecting the constant coefficients independent of $\epsilon$ and $\Delta E$, we get
	\begin{equation}
		\begin{split}
			&\frac{1}{\epsilon}\Delta E^{\frac{1}{1+\lambda}} 
			\sim
			\left(\frac{\Delta E}{\epsilon}\right)^\frac{1}{2-\tau}
			\\
			&\Delta E^{\frac{1}{2-\tau}-\frac{1}{1+\lambda}}
			\sim
			\epsilon^{\frac{1}{2-\tau}-1}
			\\
			&\Delta E^{\frac{\lambda+\tau-1}{(1+\tau){(2-\tau)}}}
			\sim
			\epsilon^{\frac{\tau-1}{2-\tau}}
			\\
			&\Delta E
			\sim
			\epsilon^{\frac{(\tau-1)(\lambda+1)}{\lambda+\tau-1}}
		\end{split}
	\end{equation}
	This concludes the derivation of Eq. (7) in the manuscript.